\documentclass[aps,prl,showpacs,amssymb,amsmath,superscriptaddress,twocolumn]{revtex4-1}
\usepackage{graphicx}
\begin{document}

\title{Non-adiabatic time-dependent density functional theory of the impurity resistivity of metals}
\author{V.~U.~Nazarov}

\affiliation{Research Center for Applied Sciences, Academia Sinica,
 Taipei 11529, Taiwan}

\author{G.~Vignale}
\affiliation {Department of Physics and Astronomy, University of Missouri, Columbia, Missouri 65211, USA}

\author{Y.-C.~Chang}
\affiliation{Research Center for Applied Sciences, Academia Sinica,
 Taipei 11529, Taiwan}

\begin{abstract}
We make use of the time-dependent density functional theory to derive a new formally exact expression for the dc resistivity of metals with  impurities.
This expression takes  fully into account the dynamics of electron-electron interactions.
Correction to the conventional $T$-matrix (phase-shifts) theory is treated within hydrodynamics of inhomogeneous viscous  electron liquid.
As a first application, we present calculations of the residual resistivity of aluminum as a function of the atomic number of the impurities.
We show that the inclusion of many-body corrections considerably improves the agreement between theory and experiment.
\end{abstract}


\maketitle

Scattering of carriers by impurities is one of the fundamental mechanisms of resistivity in solids ~\cite{Mott-58}.
Within the single-particle theory, this problem can be efficiently  addressed with the use of well established techniques
of potential scattering theory.
Electron-electron interactions can be included, at this level, within the ground-state density-functional theory \cite{Hohenberg-64,Kohn-65}
as electrostatic and exchange-correlation effective static potential which scatters the electrons as single particles~\cite{Nieminen-80,Puska-83}.
However, the single-particle approach fails to account for the {\it dynamical} exchange and correlation effects
which cannot be forced into the mold of a static mean field theory.

A powerful theoretical tool has been devised to account, in principle exactly,
for dynamical electron-electron interaction effects in inhomogeneous systems.
This is known as the time-dependent density functional theory (TDDFT)~\cite{Gross-85}.
Historically, TDDFT was developed to improve the calculation of atomic scattering cross-sections and excitation energies
in both bounded~\cite{Petersilka-96} and extended systems~\cite{Reining-02,Marini-03}.
However, in recent years this theory has often been applied to the treatment of static phenomena
(e.g., the polarizability of polymer chains~\cite{Faasen-02,Faasen-03}),  and steady-state transport phenomena
(e.g., the stopping power of metals for slow ions~\cite{Nazarov-05,Nazarov-07}, and the conductance of quantum point
contacts~\cite{Sai-05,Jung-07,Sai-07,Koentopp-06,Bokes-07}).
In these applications one needs to take the zero-frequency limit of  a time-dependent process, which can be properly described by TDDFT or
its current generalization -- the time-dependent current density functional theory (TDCDFT)~\cite{Ghosh-88,Vignale-96}.

Moving in the same direction,
in this Letter we present the first complete TDDFT formulation for the impurity resistivity of metals.
We derive an exact formula for the frequency-dependent resistivity
in terms of quantities that can be  calculated entirely within density functional theory.
While the standard Kubo formula gives a formally exact expression for the {\it conductivity},
our formula gives an expression for the resistivity, and furthermore does not require that we calculate explicitly the current distribution.
A major advantage of working with the resistivity rather than with the conductivity is that physically distinct
dissipative processes enter the resistivity as additive contributions (Matthiessen's rule~\cite{Ashcroft&Mermin}).
In particular, we find that our expression for the resistivity naturally separates  into a single-particle contribution and a dynamical
many-body contribution.  The former is shown to reduce to the classical  potential-scattering formula for the resistivity; the latter
takes into account the viscosity of the electron liquid~\cite{Conti-97}.

Our formula appears to be a promising tool for a systematic improvement on the existing calculations of the resistivity of metals.
We demonstrate this in a concrete application, namely the model calculation of the resistivity of aluminum in the presence of random impurities.
We show that  dynamical corrections, calculated with the help of the available formulas for the visco-elastic
constants of the uniform electron liquid~\cite{Conti-99},  can considerably improve the agreement between the calculated and the measured resistivity.

We start by writing down the classical (single-particle)
formula~\cite{Mott-58} for the resistivity of an electron gas of density
$\bar n_0$
with impurities randomly distributed with density $n_i$:
\begin{eqnarray}
\rho = k_F n_i \sigma_{tr}(k_F)/(e \bar n_0).
\label{rho1}
\end{eqnarray}
Here $k_F$ is the Fermi wave-vector of the electron gas, $\sigma_{tr}(k_F)$ is the transport
cross-section of an electron at the Fermi level scattered by the potential of an individual impurity,
and $e$ is the absolute value of the electron charge. The basic assumptions underlying Eq.~(\ref{rho1}) are:
(i) Electrons do not interact with each other while being scattered by the impurities,
and (ii) Electrons feel only one impurity at a time, i.e., the coherent scattering
of an electron from more than one impurity is neglected.   Both assumptions will be relaxed in the treatment that follows.

Let us consider a monochromatic and uniform external electric field
${\bf E}_{ext}(t)= {\bf E}_{ext} \, e^{-i  \omega t}$
applied to electron gas with impurities positioned at ${\bf R}_k$, $k=1,2,...$.
We can write the current density averaged over the normalization  volume $V$ as
\footnote{In Eq.~(\ref{jj}), the integration over ${\bf r}'$ gives the microscopic current-density by virtue of the Kubo formula,
while the integration over ${\bf r}$ with the division by $V$ represents the averaging in
the system uniform on the macroscopic scale. All  many-body and possible quantum interference effects are contained in the current-density
response function $\hat{\chi}$.}
\begin{eqnarray}
j_i(\omega)= \frac{i c }{ \omega V}
\int\limits_{V} d {\bf r} d {\bf r}' \hat{\chi}_{ij}({\bf r},{\bf r}',\omega) E_{ext,j},
\label{jj}
\end{eqnarray}
where $\hat{\chi}_{ij}({\bf r},{\bf r}',\omega)$ is the current-density
response function of the inhomogeneous electron gas with  impurities.
A summation over the repeated Cartesian index $j$ is implied.
We transform Eq.~(\ref{jj})
with the help of the sum-rule
\footnote{Equation (\ref{SRS}) is a generalization for infinite systems of the corresponding
sum-rule of Ref.~\cite{Vignale-B} taking into account the
potential
of the positive background  $V_b(r)=2\pi \bar n_0 \,e^2  r^2$,
which, as well as the bare potential of the impurities $V_0(r)$, is external to the electron subsystem.}
\begin{eqnarray}
&&c \left( \omega^2- \omega_p^2\right) \int
\hat{\chi}_{ij}({\bf r},{\bf r}',\omega) \, d {\bf r}'= \frac{c}{m} \times \cr\cr
&& \int
\hat{\chi}_{ik}({\bf r},{\bf r}',\omega) \, \nabla'_k \nabla'_j V_0({\bf r}')
\, d {\bf r}' + \frac{e \, \omega^2}{m} n_0({\bf r})\,\delta_{ij},
\label{SRS}
\end{eqnarray}
where
\begin{eqnarray}
V_0({\bf r})=\sum\limits_k v_0({\bf r}-{\bf R}_k),
\end{eqnarray}
$v_0({\bf r})$ is the {\em bare} potential of one impurity centered at origin,
$n_0({\bf r})$ is the ground-state electron density,
$
\omega_p=\sqrt{4 \pi e^2 \bar n_0/m}
$
is plasma frequency of the homogeneous electron gas {\em without} impurities,
and $c$ and $m$ are the speed of light in vacuum and the mass of electron, respectively.
Applying Eq.~(\ref{SRS}) twice with respect to the integration over ${\bf r}$ and ${\bf r}'$ in Eq.~(\ref{jj}), and
using the expression for the  density-response function
\begin{eqnarray*}
\chi({\bf r},{\bf r}',\omega) =  -\frac{c}{e \, \omega^2} \nabla_i \cdot \hat{\chi}_{ij}({\bf r},{\bf r}',\omega)
\cdot  \nabla'_j,
\end{eqnarray*}
together with the static sum-rule \cite{Nazarov-05}
\begin{eqnarray}
\int  \chi ({\bf r},{\bf r}',0 )  \nabla'_i V_0({\bf r}') \, d{\bf r}' = \nabla_i n_0({\bf r}),
\label{SSR}
\end{eqnarray}
we eventually write the current-density as
\begin{eqnarray}
j_i(\omega) \! = \! \frac{i e \omega}{ m  (\omega^2 \!- \!\omega_p^2)} \!
\left\{  \!  \bar n_0  E_{ext,i} \! + \! \frac{1}{m  (\omega^2 \! - \! \omega_p^2) V} \! \left[ \! \int\limits_V \! d{\bf r} \! \int\limits_V \! d{\bf r}' \right. \right. \cr\cr
\left. \left.
   \times [\nabla_i V_0(r)] [\chi ({\bf r},{\bf r}',\omega)  \! - \! \chi ({\bf r},{\bf r}',0)]  [\nabla'_j V_0(r')]
\right] \!  E_{ext,j}
\right\}.
\label{jjj}
\end{eqnarray}

Independently of the foregoing considerations, just from Maxwell's equations for the electromagnetic field,   we find that the resistivity can be written as 
\footnote{See EPAPS document No. [ ], Secs. I, II, III, and IV,
for further details on Eq.~(\ref{rhop}), the derivation of Eqs.~(\ref{rho12})-(\ref{rho2}) from Eq.~(\ref{rhopp}),
the equivalence between Eqs.~(\ref{rho1}) and (\ref{rho10}),
and on the phase-shifts based DFT evaluation of $\rho_1$ from Eq. (\ref{rho1}), respectively.}
\begin{eqnarray}
\rho(\omega)=\frac{E_i(\omega)}{ j_i(\omega)}=\frac{1}{\omega} \left(4 \pi i e+ \frac{\omega E_{ext}^2}{j_{i} E_{ext,i}} \right).
\label{rhop}
\end{eqnarray}

From Eq.~(\ref{jjj}) we conclude  that the expression in the parentheses of Eq.~(\ref{rhop}) is zero at $\omega=0$.
Taking the limit $\omega\rightarrow 0$ in Eq.~(\ref{rhop}) by   L'Hopital's rule  and using Eq.~(\ref{jjj}) again, we arrive at
\begin{eqnarray}
\rho = && - \frac{1}{e \bar n_0^2 V}
 \int\limits_V
[\nabla V_0({\bf r})\cdot \hat{{\bf E}}_{ext}] [\nabla' V_0({\bf r}') \cdot \hat{{\bf E}}_{ext}]
\cr\cr && \times \frac{\partial {\rm Im} \chi({\bf r},{\bf r}',\omega)}{\partial \omega}\Bigr |_{\omega=0} d {\bf r} \, d {\bf r}',
\label{rhopp}
\end{eqnarray}
where $\hat{{\bf E}}_{ext}$ is the unit vector parallel to ${\bf E}_{ext}$.

Equation (\ref{rhopp}) is the formal solution to the problem of expressing the resistivity in terms of the
density-density response function $\chi$ of the {\em interacting} inhomogeneous electron gas with impurities.
Using the relation \cite{Gross-85}
\begin{eqnarray}
\chi^{-1}({\bf r},{\bf r}',{\omega}) = \chi^{-1}_{KS}({\bf r},{\bf
r}',{\omega})\! - \! \! f_{xc}({\bf r},{\bf r}',{\omega}) - \!
\frac{1}{|{\bf r}-{\bf r}'|}, \label{fs}
\end{eqnarray}
we can conveniently rewrite Eq.~(\ref{rhopp}) in terms of the Kohn-Sham (KS)
density-density response function $\chi_{KS}$ of {\em non-interacting} electrons  and the dynamical exchange and correlation kernel $f_{xc}$
\footnotemark[\value{footnote}]
 \begin{eqnarray}
\rho=\rho_1+\rho_2,
\label{rho12}
\end{eqnarray}
\begin{eqnarray}
\rho_1 = && - \frac{1}{e \bar n_0^2 V}
 \int\limits_V
[\nabla V_{KS}({\bf r})\cdot \hat{{\bf E}}_{ext}] [\nabla' V_{KS}({\bf r}') \cdot \hat{{\bf E}}_{ext}]
\cr\cr && \times \frac{\partial {\rm Im} \chi_{KS}({\bf r},{\bf r}',\omega)}{\partial \omega}\Bigr |_{\omega=0} d {\bf r} \, d {\bf r}',
\label{rho10}
\end{eqnarray}
where $V_{KS}({\bf r})$ is the static KS  potential, and
\begin{eqnarray}
\rho_2= && -\frac{1}{e \bar n_0^2 V} \int\limits_V
[\nabla_{\bf r} n_0({\bf r})\cdot \hat{{\bf E}}_{ext}]  [\nabla_{{\bf r}'} n_0({\bf r}')\cdot \hat{{\bf E}}_{ext}] \cr\cr
&& \times \frac{\partial \,{\rm Im} f_{xc}({\bf r},{\bf r}',\omega)}{\partial\omega}
\Bigr |_{\omega=0}
\, d{\bf r}  \, d{\bf r}'.
\label{rho2}
\end{eqnarray}

The first equation~(\ref{rho10}) is the single-particle (KS) contribution to the resistivity.  The second (\ref{rho2})  is the dynamical exchange-correlations contribution.
If the frequency dependence of $f_{xc}$ is neglected -- as one does, for example, in the adiabatic approximation to TDDFT  --  then  Eq.~(\ref{rho2}) yields $\rho_2=0$.

To establish the connection between $\rho_1$ and the classical potential-scattering result of Eq.~(\ref{rho1}), we must neglect in Eq.~(\ref{rho10}) the coherent scattering from multiple impurities. To do this we replace
the full  KS potential $V_{KS}({\bf r})$ by the KS potential associated with a single impurity in the electron gas,  and we interpret the KS response function $\chi_{KS}({\bf r},{\bf r}',\omega)$ accordingly. The normalization volume is taken to be equal to the volume per impurity, i.e., $V=1/n_i$.
It can be rigourously proved \footnotemark[\value{footnote}] that Eq.~(\ref{rho10}),  thus modified,  is equivalent to Eq.~(\ref{rho1}).
This result, combined with the discussion of the previous paragraph,
leads to the important conclusion that the adiabatic approximation to TDDFT is equivalent to the classical potential-scattering
($T$-matrix) approach as far as the calculation of the resistivity is concerned.

The single-particle contribution to the resistivity is conventionally obtained from Eq.~(\ref{rho1})
\footnotemark[\value{footnote}],
using the $T$-matrix (phase-shift) technique to calculate the scattering cross-section from the static KS potential~\cite{Nieminen-80,Puska-83}.   To find the many-body contribution to the resistivity from Eq.~(\ref{rho2}), we need a good approximation
to the dynamical exchange and correlation kernel $f_{xc}$.  It is known~\cite{Vignale-95} that  $f_{xc}({\bf r},{\bf r'})$ is strongly non-local (i.e. a long-ranged function of  $|{\bf r}-{\bf r'}|$) and this non-locality is crucial to a proper description of many-body effects in transport phenomena, even on a qualitative level
\cite{Nazarov-07}.
This immediately poses the problem of constructing a reasonably accurate non-local approximation for $f_{xc}$.  In a recent paper~\cite{Nazarov-07} we have shown how this can be done starting from an exact representation of the scalar $f_{xc}$ kernel in terms of the tensorial exchange and correlation kernel $\hat{f}_{xc}$ of time-dependent {\it current} density functional theory.  This representation reads
\begin{eqnarray}
&&\!\!\! \! f_{xc} \! \! = \!
-\frac{e \, \omega^2}{c}
\nabla^{-2} \nabla \! \cdot \!
\left\{ \!
\hat{f}_{xc}
\! \! + \! \! \left(\hat{\chi}^{-1}_{KS} \! \!-\!\! \hat{f}_{xc}\right)\!
\left[\! \hat{T} \! \left(\hat{\chi}^{-1}_{KS} \!-\!\! \hat{f}_{xc} \! \right)\! \hat{T} \right]^{-1} \right. \cr\cr
&&\left.  \times \left(\hat{\chi}^{-1}_{KS} \!-\! \! \hat{f}_{xc}\right)\!-\! \hat{\chi}^{-1}_{KS}
\left(\hat{T}\hat{\chi}^{-1}_{KS}\hat{T}\right)^{-1}\!\!\!\!
\hat{\chi}^{-1}_{KS}
\right\} \cdot  \nabla \nabla^{-2},
\label{via}
\end{eqnarray}
where $\hat{\chi}_{KS}$ is the KS current-density response function and $\hat{T}$ is the projector operator onto the subspace of transverse vector fields (i.e. divergence-free fields)
\footnote{In the case of the one-dimensional inhomogeneity, Eq.~(\ref{via}) simplifies to an expression which does not involve $\hat{\chi}_{KS}$ \cite{Vignale-B,Dion-05}.
This, however, is not true in the general case.}.
By making use of the local density approximation (LDA) for the tensorial   $\hat{f}_{xc}$ in the right-hand side of Eq.~(\ref{via}),
we obtain a non-local approximation for  the scalar $f_{xc}$, which satisfies the zero-force sum-rule  requirements
[\onlinecite{Vignale-95},\onlinecite{Nazarov-07}],
and can, therefore, be considered a promisingly accurate approximation for transport problems.

\begin{figure}[h]
\includegraphics[width=0.475\textwidth,height=0.35\textwidth]{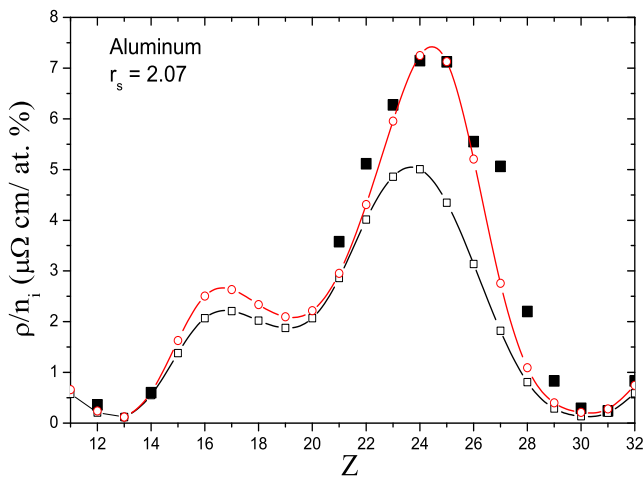}
\caption{\label{Fig}(Color online) Residual resistivity of aluminum due to substitutional impurities
of atomic number $Z$, as a function of $Z$. The chained curve with circles (red online) is our result
with inclusion of the dynamical exchange and correlations (the sum of $\rho_1$ and $\rho_2$ obtained with use of Eqs.~(\ref{rho1}) and (\ref{rho2}), respectively). The chained curve with squares (black online) is the result
of the single-particle theory ($\rho_1$ only).
Solid squares
are experimental data compiled from Refs. \cite{Babic}.
} \label{Fig1}
\end{figure}

In Refs.~\cite{Vignale-96,Vignale-97}, the LDA to the  exchange and correlation kernel $\hat{f}_{xc}$ of the TDCDFT has been worked out within the framework of the hydrodynamics of inhomogeneous viscous electron liquid.
We, therefore, use Eq.~(\ref{rho2}) with $f_{xc}$ given
by Eq.~(\ref{via}) and $\hat{f}_{xc}$ as expressed in Ref.~\cite{Vignale-97} through the viscoelastic constants of electron liquid.
In Fig.~\ref{Fig}, we present results for resistivity 
for substitutional impurities of atomic number $Z$ from 11 through 32 in an aluminum host. The latter is modeled as a jellium with Wigner-Seitz radius $r_s=2.07$.
In this calculation we have neglected the coherent scattering from multiple impurities, focusing instead  on the many-body dynamical exchange and correlations effects. The values of the viscoelastic constants  were taken from Ref.~\cite{Conti-99}.  Our purpose is not to take into account all the effects that could possibly contribute to the resistivity in a real solid aluminum, but rather to show that the many-body viscosity corrections are sizeable and indeed of the right order of magnitude to account for the observed discrepancy between available theoretical calculations and experimental data.

The single-particle contribution $\rho_1$ calculated from Eq.~(\ref{rho1}) and represented by the chained curve with squares (black online)
is found to be in agreement with earlier calculations \cite{Nieminen-80,Puska-83}.
The total resistivity,  including dynamical exchange and correlation contributions, is represented by the chained curve with circles
(red online).  An improved agreement between theory and experiment  can be clearly seen from the figure.
The effects left out by our calculation that could possibly contribute to the remaining disagreement  between the theory and experiment are
(i) the band structure and lattice distortion effects, (ii) the possible spin-polarization, and (iii) the coherent scattering by the  impurities at different sites.
Another potentially important source of error can be in the values of the visco-elastic constants of the electron liquid.

Finally we note that Eq.~(\ref{rhopp}) allows us to establish a general relation between the impurity resistivity  and the friction coefficient
\footnote{The friction coefficient of a medium for an atom is the ratio of the stopping power of this material for the atom and the atom's velocity, taken at the zero-velocity value.} of the same host for the same type of impurity atom. The latter can be written as \cite{Nazarov-05}
\begin{eqnarray}
Q = &&
 - \int
[\nabla V_0(r)\cdot \hat{{\bf v}}] [\nabla' V_0(r') \cdot \hat{{\bf v}}]
\cr\cr && \times \frac{\partial {\rm Im} \chi({\bf r},{\bf r}',\omega)}{\partial \omega}\Bigr |_{\omega=0} d {\bf r} \, d {\bf r}',
\label{Q}
\end{eqnarray}
where ${\bf v}$ is the velocity of the atom, and, comparing with Eq.~(\ref{rhopp}), we can write
\begin{equation}
\rho=n_i Q / (e \bar n_0^2).
\label{rel}
\end{equation}
We point out that the relation (\ref{rel}) quite generally holds {\em within the many-body theory}
and is a stronger statement than
$
\rho_1=n_i Q_1 / (e \bar n_0^2),
$
which is a simple consequence of Eq.~(\ref{rho1}) and the corresponding single-particle result for the friction coefficient
$
Q_1= \bar n_0  \, k_F \sigma_{tr}(k_F)
$
\cite{Finneman-68}.

In conclusion, we have developed the non-adiabatic
time-dependent density functional formalism for a systematic calculation of the dc residual resistivity of metals with  impurities.
The contribution to the resistivity arising from the many-body interactions
has been expressed through the dynamical exchange and correlation kernel $f_{xc}$.
We have shown that all the dynamical effects of the  electron-electron interaction are contained in the frequency dependence of $f_{xc}$.
The adiabatic approximation, which neglects this frequency dependence,  is  exactly equivalent to the conventional single-particle
potential-scattering theory of the resistivity, provided the coherent scattering from multiple impurities is neglected as well.
Our calculations of the residual resistivity of Al with various impurity atoms of different nuclear charge show that the inclusion
of dynamical exchange and correlation  considerably improves the agreement between theory and experiment.

\acknowledgments
VUN  acknowledges support from National Science Council, Taiwan, Grant No. 100-2112-M-001-025-MY3.
GV acknowledges support from DOE Grant No. DEFG02-05ER46203.

\end{document}